\documentclass[aps,prx,twocolumn,linenumbers,showpacs,floatfix,linenumbers,superscriptaddress,nofootinbib]{revtex4}

\usepackage{times}
\usepackage{physics}
\usepackage{graphicx}
\usepackage{hyperref}
\hypersetup{colorlinks=true,linkcolor=blue,citecolor=blue}
\usepackage{amsmath}
\usepackage{wrapfig}
\usepackage{siunitx}
\usepackage{upgreek}
\bibliography{Sci,books,arXiv}

\begin{document}

\title{Acoustic ferromagnetic resonance and spin pumping induced by surface acoustic waves}

\begin{abstract}
Voltage induced magnetization dynamics of magnetic thin films is a valuable tool to study anisotropic fields, exchange couplings, magnetization damping and spin pumping mechanism. A particularly well established technique is the ferromagnetic resonance (FMR) generated by the coupling of microwave photons and magnetization eigenmodes in the GHz range. Here we review the basic concepts of the so-called acoustic ferromagnetic resonance technique (a-FMR) induced by the coupling of surface acoustic waves (SAW) and magnetization of thin films. Interestingly, additional to the benefits of the microwave excited FMR technique, the coupling between SAW and magnetization also offers fertile ground to study magnon-phonon and spin rotation couplings. We describe the in-plane magnetic field angle dependence of the a-FMR by measuring the absorption / transmission of SAW and the attenuation of SAW in the presence of rotational motion of the lattice, and show the consequent generation of spin current by acoustic spin pumping.     
\end{abstract}
\date{\today}

\author{Jorge Puebla}
\email{jorgeluis.pueblanunez@riken.jp}
\affiliation{CEMS, RIKEN, Saitama, 351-0198, Japan}
\author{Mingran Xu}
\affiliation{CEMS, RIKEN, Saitama, 351-0198, Japan}
\affiliation{Institute for Solid State Physics, University of Tokyo, 5-1-5 Kashiwanoha, Kashiwa, Chiba, 277-8581, Japan}
\author{Bivas Rana}
\affiliation{CEMS, RIKEN, Saitama, 351-0198, Japan}
\author{Kei Yamamoto}
\affiliation{CEMS, RIKEN, Saitama, 351-0198, Japan}
\affiliation{Advanced Science Research Center, Japan Atomic Energy Agency, Tokai 319-1195, Japan}
\author{Sadamichi Maekawa}
\affiliation{CEMS, RIKEN, Saitama, 351-0198, Japan}
\affiliation{Advanced Science Research Center, Japan Atomic Energy Agency, Tokai 319-1195, Japan}
\affiliation{Kavli Institute for Theoretical Sciences, University of Chinese Academy of Sciences, Beijing 100049, People’s Republic of China}
\author{Yoshichika Otani}
\email{yotani@issp.u-tokyo.ac.jp}
\affiliation{CEMS, RIKEN, Saitama, 351-0198, Japan}
\affiliation{Institute for Solid State Physics, University of Tokyo, 5-1-5 Kashiwanoha, Kashiwa, Chiba, 277-8581, Japan}
\maketitle

\section{introduction}

Arguably, one of the most enlightening works by the late physicist Charles Kittel\footnote{Charles Kittel passed away the last 15th of May 2019 at the age of 102} is the theoretical description of the ferromagnetic resonance absorption, published first in 1947 \cite{Kittel47} and extended to shape anistropies in 1948 \cite{Kittel48}. At these early works, Kittel described the magnetization dynamics exerted in a ferromagnetic specimen subjected to a strong dc field $H_{z}$ and a weak perpendicular microwave field $H_{x}$, such that the magnetization dynamics can be well described by $\partial\boldsymbol{M}/\partial t=\upgamma[\boldsymbol{M}\times\boldsymbol{H}]$; where $\upgamma$ is the gyromagnetic ratio, $\boldsymbol{M}$ the magnetization and $\boldsymbol{H}$ is the external field with components $(H_{x}, H_{y}, H_{z})$. If the dc field is in-plane and strong enough to fully align the magnetization $\boldsymbol{M}$, then the magnetization precess as a single magnetic domain, a phenomenon we know nowadays as ferromagnetic resonance (FMR). The corresponding frequency at the resonance condition $f_{0}$ is described by the so-called Kittel formula $f_{0}=\frac{\upgamma}{2\pi}\sqrt{H_z ( H_z +\upmu_{0} \abs{\boldsymbol{M}})}$. 

Nowadays, the FMR is a versatile tool that allows studying magnetization dynamics in thin films \cite{Bilzera}, spin waves \cite{Boone}, magnetization switching \cite{Thirion} and spin pumping \cite{Rojas}. Remarkably, 10 years after his description of FMR, it was the same Charles Kittel who first formulated the coupling of spin waves (magnons) and lattice vibrations (phonons) at resonance conditions, giving origin to the acoustic excited FMR \cite{Kittel58}. One initial conclusion was that microwave phonons were necessary for reaching the resonance condition. Conveniently, microwave phonons can be generated by interdigital transducers fabricated on top of piezoelectric substrates. Additional to the previous description of applications of standard FMR (microwave field excited), Kittel suggested that magnetic bulk crystals may show nonreciprocal acoustic properties, and induce strong phonon attenuation. 

Here, we first overview the main characteristics of acoustic ferromagnetic resonance (a-FMR) excited by GHz frequency surface acoustic waves (SAW). Then, we review the description of the magnetization coupling with elastic rotation and its dependence when varying in-plane magnetic field \cite{Mingran, Dreher}. Furthermore, we recalled an example of the generation of spin current by a-FMR, the so-called acoustic spin pumping. We show that the order of magnitude of the spin current density is of the same order of the more standard spin pumping mechanism excited by microwave photons. Additional to the most recent works of a-FMR \cite{Mingran, Dreher}, we describe the less explored coupling of magnetization with lattice rotation, first published more than 40 years ago by one of the authors of the present review \cite{Maekawa}. Finally, beyond the a-FMR study we conclude with an outlook for the coupling of SAW with magnetic and nonmagnetic materials for spintronic research and applications.       

\section{Acoustic ferromagnetic resonance}

\subsection{Surface acoustic waves}

As its name suggests, surface acoustic waves (SAW) are elastic waves that travel parallel to the surface of an elastic material, with a decay perpendicular to the surface into the bulk with an approximated decay length equal to the acoustic wavelength $\uplambda_{SAW}$. Figure 1(a) shows the schematics of a two port interdigital transducer (IDT) device on a piezoelectric substrate (LiNbO$_{3}$, Lithium Niobate) for generation of SAW, the wavelength of acoustic waves, $\uplambda_{SAW}$, depends on the periodicity of IDTs as also shown in the schematics. The generation of SAW is done by the inverse piezoelectric effect, where elastic deformation is produced by injection of rf-voltage. The reading of SAW is done by the direct piezoelectric effect, converting elastic deformation back to voltage signal. The SAW frequency is defined by $f_{SAW}=\frac{v_{s}}{\uplambda_{SAW}}$, where $v_{s}$ is the sound velocity of the piezoelectric material. Figure 1(b) shows a scanning electron microscope (SEM) image of IDTs with 400 nm width and $\uplambda_{SAW}=1.6$~$\upmu$m. The IDTs are patterned by electron beam lithography and made of Ti(5nm)/Au(30nm) by the lift-off method. Since the sound velocity $v_{s}=4000$~m/s in LiNbO$_{3}$, we expect the generation of SAW with an approximate frequency $f_{SAW}=2.5$ GHz. Figure 1(c) shows the characterization of the scattering parameter for transmission S$_{12}$ by a vector network analyzer (VNA), giving a SAW resonance frequency $f_{SAW}=2.38$ GHz. The voltage generation of SAW also induces generation of spurious electromagnetic waves (EMW); since the SAW and EMW velocities are different, it is possible to filter out the EMW by a technique called time-gating that employs Fourier transform operations. IDTs act not only as the generator and receiver for acoustic phonons, but also as an antenna, receiving the microwave signals through the air. And acoustic phonon is propagating in a speed of $v_{s}$, which is in thousand meters per second level, while the velocity of EMW is approximately 3$\times$10$^{8}$ m/s. Hence, SAW signal arrive to the second IDT port after the EMW. As an additional function of our VNA, we are able to analyze the signal in time-domain. In order to rule out the noise brought by the EMW, we set a gate in time-domain, to take the signal which arrives in the interval expected according the the SAW velocity $v_{s}$ and the lenght of our device. And by using Fourier transform, we obtain frequency-domain signal. The comparision of the signal with and without time gating is presented in figure 1(c).

\begin{figure}[t!]
\begin{center}
\includegraphics[width=1.0\columnwidth, height=10.0cm, keepaspectratio]{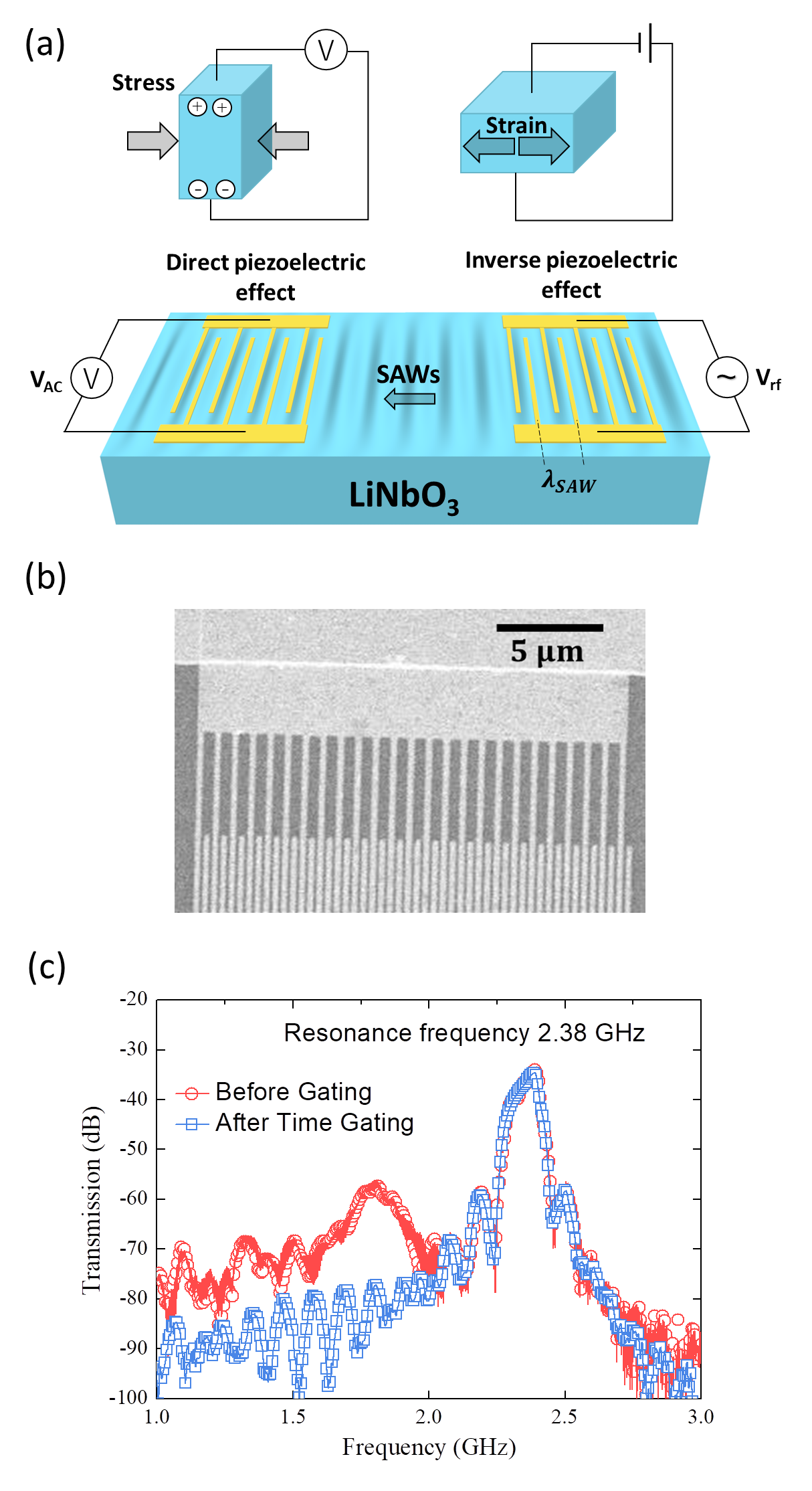}
\end{center}
\vspace*{-5mm}
\caption{(a) Illustration of experiment setup. The surface acoustic waves (SAW) are generated by applying RF voltage on an interdigital transducer port (IDT) via the inverse piezoelectric effect. A second IDT port measures the transmittance of SAW via the direct piezoelectric effect. (b) Scanning electron microscope (SEM) image of IDT fingers with a nominal width of 400 nm. (c) Transmittance measurement of SAW by vector network analyzer, the measurement shows a SAW resonance frequency, $f_{SAW}=2.38$ GHz. We show the measurement before (red open circles) and after (blue open squares) time gating filtering removal of electromagnetic waves.}
\label{fig:SAW}
\end{figure}

\subsection{Magnetization coupling with elastic strain}

When SAWs propagate on a ferromagnetic layer it produces a time variant strain tensor field $\hat{\varepsilon }(t)$ in the lattice, which couples to the local magnetic environment via the magnetoelastic effect; as a consequence a time varying magnetic field is exerted. Here, we describe how this time varying magnon-phonon coupling induces a-FMR.  

\begin{figure}[t!]
\begin{center}
\includegraphics[width=1.0\columnwidth, height=10.0cm, keepaspectratio]{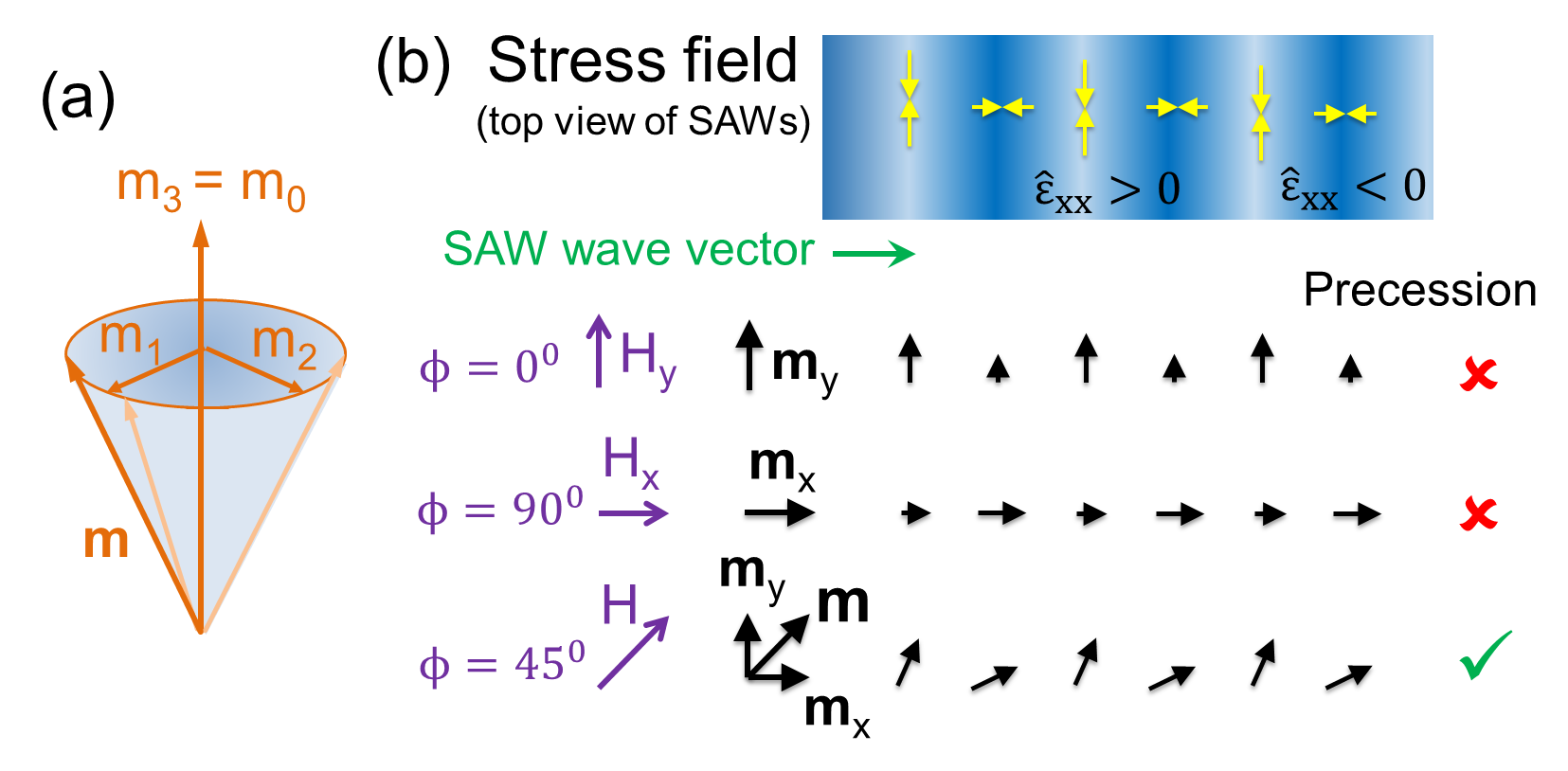}
\end{center}
\vspace*{-5mm}
\caption{(a) Schematic representation of the coordinates of magnetization components $m_{1}$, $m_{2}$ and $m_{3}$ ($m_{0}$). (b) Schematic representation of stress fields of SAW from a top view. The maximum magnetoelastic torque is induced at $\upphi=45^{\circ}$, and vanishes at $\upphi=0^{\circ}$ and $\upphi=90^{\circ}$}
\label{fig:FMR}
\end{figure}

\begin{figure}[t!]
\begin{center}
\includegraphics[width=1.0\columnwidth, height=10.0cm, keepaspectratio]{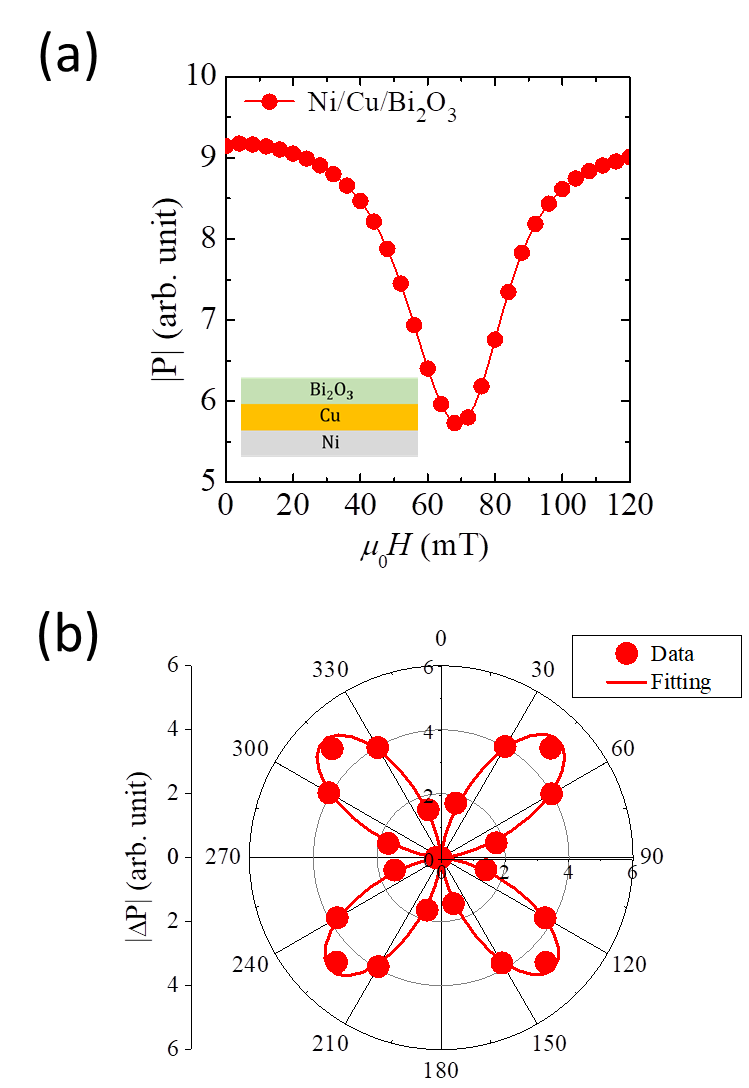}
\end{center}
\vspace*{-5mm}
\caption{(a) Power absorption $P_{SAW}$ of SAW at resonance condition for FMR driving of a Ni layer of 10nm within a Ni/Cu/Bi$_{2}$O$_{3}$ heterostructure (inset). (b) Normalized in-plane magnetic field angle dependence of $\Delta P_{SAW}$. Figure (b) is adapted from ref.\cite{Mingran}.}
\label{fig:ABS1}
\end{figure}

Let us start by describing first the dynamics exerted on the normalized magnetization vector $\boldsymbol{m}=\boldsymbol{M}/ \abs{ \boldsymbol{M}}  $ when an effective driving magnetic field $\upmu_{0}\boldsymbol{H}_{\rm eff}$ is present. Such dynamics are described by the Landau-Lifshitz-Gilbert (LLG) equation
\begin{equation} 
\partial\boldsymbol{m}/\partial t=\upgamma\boldsymbol{m}\times\upmu_{0}\boldsymbol{H}_{\rm eff}
+\upalpha\boldsymbol{m}\times\partial\boldsymbol{m}/\partial t
\end{equation}
where $\upalpha$ is the Gilbert damping constant and the gyromagnetic ratio $\upgamma$ is taken as negative. In standard FMR the $\upmu_{0}\boldsymbol{H}_{\rm eff}$ is given by the magnetic field generated by microwave photons, commonly achieved by coplanar waveguides. In a SAW excited a-FMR, the magnetoelastic coupling links the time dependent lattice strain tensor $\hat{\varepsilon}(t)$ to the magnetic environment. To calculate $\upmu_{0}\boldsymbol{H}_{\rm eff}$ as a function of the strain tensor components $\varepsilon _{ij}$ and magnetization components $m_i $ where $i,j = x,y,z$; we phenomenologically postulate the thermodynamic relation of Gibbs $\upmu_{0}\boldsymbol{H}_{\rm eff}=-\nabla_{m}G^{tot}$, which should be valid for linear response around equilibrium: Follwing Dreher et al.~\cite{Dreher}, we have introduced $G^{tot}=G+G_{me}$ where $G$ is the free energy density normalized by $\abs{\boldsymbol{M}}$ in a ferromagnet, which may for instance be taken to be
\begin{equation} 
G=-\upmu_{0}\boldsymbol{H}\cdot \boldsymbol{m}+B_{d}m_{z}^{2}+B_{u}(\boldsymbol{m} \cdot \boldsymbol{u})^{2}
\label{eq:Gibbs} .
\end{equation}     
Here $B_{d} = \upmu _0 \abs{\boldsymbol{M}} /2 $ is the shape anisotropy and $B_{u}$ is the in-plane uniaxial anisotropy along the unit vector $\boldsymbol{u}$. Although one could further include other terms such as dipole-dipole and exchange interactions, the detailed form of $G$ is largely irrelevant in the following discussions and we omit them here. $G_{me}$ is the magnetoelasitc contribution to the ($\abs{\boldsymbol{M}}$-normalized) free energy density, which for a cubic crystal reads 
\begin{align} 
G_{me}  =&b_{1} ( \varepsilon_{xx} m_x^2 + \varepsilon _{yy}m_y^2 + \varepsilon _{zz} m_z^2  )  \nonumber \\
& + 2 b_2 ( \varepsilon _{xy} m_x m_y + \varepsilon _{yz} m_y m_z + \varepsilon _{zx} m_z m_x ) ,
\end{align} 
where $b_{1,2}$ are the magnetoelastic coupling constants. We define a new coordinate system ($x_{1}, x_{2}, x_{3}$) so that the magnetization components are accordingly ($m_{1}, m_{2}, m_{3}$); where the equilibrium direction of the magnetization $\boldsymbol{m}_{0}$ lies in $x_{3}$ ($m_{3}$), while $m_{1}$ and $m_{2}$ are the small dynamical components perpendicular to each other and perpendicular to $m_{3}$ (see schematic in fig.~\ref{fig:FMR}(a)). Now we can expand the effective field $\upmu_{0}\boldsymbol{H}_{\rm eff}$ under the influence of magnetoelastic coupling to first order in $m_{1,2}$
\begin{equation} 
\upmu_{0}\boldsymbol{H}_{\rm eff}= -\begin{pmatrix}{G_{11}m_{1}+G_{12}m_{2}}\\{G_{12}m_{1}+G_{22}m_{2}}\\{G_{3}}\end{pmatrix} + \begin{pmatrix}{\upmu_{0}h_{1}}\\{\upmu_{0}h_{2}}\\{\upmu_{0} h_{3}}\end{pmatrix} .
\end{equation} 
Here $G_3 = \partial _{m_3}G |_{\boldsymbol{m}=\boldsymbol{m}_0}$ and $G_{ab} = \partial _{m_a }\partial _{m_b}G |_{\boldsymbol{m}=\boldsymbol{m}_0} $, $a,b= 1,2$ are constants whose details are not needed. The transverse components of the strain induced field $h_{1,2}$ are given shortly while the longitudinal component $h_{3}$ is irrelevant in our present setup. Since the components of the driving field $\upmu_{0}\boldsymbol{H}_{\rm eff}$ that induce a-FMR are those transverse to the equilibrium of magnetization $\boldsymbol{m}_{0}$ ($m_{3}$), we rewrite the LLG equation in its matrix form as 
\begin{equation} 
\begin{pmatrix}G_{11}-G_{3} + \frac{i\omega\upalpha}{\upgamma}&G_{12} - \frac{i\omega}{\upgamma}\\G_{12}+ \frac{i\omega}{\upgamma}&G_{22}-G_{3}+ \frac{i\omega\upalpha}{\upgamma}\end{pmatrix}\begin{pmatrix}m_{1}\\m_{2}\end{pmatrix}=\upmu_{0}\begin{pmatrix}h_{1}\\h_{2}\end{pmatrix}
\end{equation}  
Solving for the transverse components ($m_{1}$, $m_{2}$) to the magnetization in equilibrium $\boldsymbol{m}_{0}$, we have
\begin{equation} 
\abs{\boldsymbol{M}} \begin{pmatrix}m_{1}\\m_{2}\end{pmatrix}=\upchi\begin{pmatrix}h_{1}\\h_{2}\end{pmatrix}
\end{equation} 
where $\upchi$ is the Polder susceptibility tensor which describes the dependence on the material parameters and static magnetic field following from the free energy density $G$ of Eq.~(\ref{eq:Gibbs})
\begin{equation} 
\upchi=\upmu _0 \abs{\boldsymbol{M}} \begin{pmatrix}G_{11}-G_{3}+ \frac{i\omega\upalpha}{\upgamma}&G_{12}-\frac{i\omega}{\upgamma}\\G_{12}+\frac{i\omega}{\upgamma}&G_{22}-G_{3}+\frac{i\omega\upalpha}{\upgamma}\end{pmatrix}^{-1}.
\end{equation} 
$h_{1}$ and $h_{2}$ are the transverse components of the driving field induced by elastic strain, such that 
\begin{align}
\upmu_{0}h_{1}=&-2b_{1}\sin\uptheta \cos\uptheta[\varepsilon_{xx}\cos^{2}\upphi+\varepsilon_{yy}\sin^{2}\upphi-\varepsilon_{zz}] \nonumber \\
&-2b_{2}[\cos2\uptheta(\varepsilon_{xz}\cos\upphi+\varepsilon_{yz}\sin\upphi)+2\varepsilon_{xy} \nonumber \\
&\sin\uptheta \cos\uptheta \sin\upphi \cos\upphi] , \label{eq:H-NV} \\
\upmu_{0}h_{2} =& 2b_{1}\sin\uptheta \sin\upphi \cos\upphi[\varepsilon_{xx}-\varepsilon_{yy}]-2b_{2}[\cos\uptheta \nonumber \\&(\varepsilon_{yx}\cos\upphi-\varepsilon_{xz}\sin\upphi)+\varepsilon_{xy}\sin\uptheta \cos2\upphi]
\label{eq:H-NV} .
\end{align} 
We observe that different from conventional FMR, the a-FMR has a strong dependence in the angles $\uptheta$ (out of plane angle) and $\upphi$ (in plane angle) between the magnetization $\boldsymbol{m}_{0}$ and the strain components $\varepsilon _{ij}$. Rayleigh waves in SAW contain the strain components $\varepsilon_{xx}$, $\varepsilon_{xz}$, where the dominant component is the longitudinal strain ($\varepsilon_{xx}$). For a pure longitudinal strain $\varepsilon_{xx}$ when the $\boldsymbol{m}_{0}$ is in-plane configuration $\uptheta=0^{\circ}$, the maximum value of the driving field $\upmu_{0}h_{1(2)}$ is at $\upphi=45^{\circ}$ where the magnetoelastic torque is larger, and vanishes at $\upphi=0^{\circ}$ and $\upphi=90^{\circ}$, as schematically shown in fig.~\ref{fig:FMR}(b). This angle dependence gives origin to the now characteristic four-fold butterfly shape of the a-FMR excited by SAW~\cite{Mingran, Dreher, Wieler}. While the SAW power absorption $\Delta P_{SAW}(\upphi) $ is in general proportional to the imaginary part of $( h_1^{\ast } ,h_2^{\ast }) \chi (h_1 ,h_2 )^T $~\cite{Dreher}, its angular dependence in the in-plane configuration is well-captured by the approximate formula~\cite{Mingran}
\begin{equation}
\Delta P_{SAW}(\upphi ) \propto [b_{1}\overline{\varepsilon_{xx}}\sin\upphi \cos\upphi \mp 2b_{2}\overline{\varepsilon_{xz}}\cos \upphi]^{2}
\label{eq:power} , 
\end{equation}
where $\overline{\varepsilon _{xx,xz}}$ are the amplitudes of $\varepsilon _{xx,xz}$ respectively and $\mp $ corresponds to the SAW propagation along $\pm \hat{\boldsymbol{x}}$ directions.
Figure \ref{fig:ABS1}(a) shows the spectrum at resonance condition for FMR driving of a Ni layer of 10nm within a Ni/Cu/Bi$_{2}$O$_{3}$ heterostructure (inset), with an external magnetic field angle $\upphi=240^{\circ}$. The full in-plane magnetic field angle dependence of the SAW power absorption $\Delta P_{SAW}$ shows the four-fold butterfly shape of a-FMR in figure \ref{fig:ABS1}(b) which is well-fitted by Eq.~(\ref{eq:power}).

\subsection{Possible role of magnetization coupling with lattice rotation}

The four-fold butterfly shape is expected to be symmetric for pure longitudinal strain $\varepsilon_{xx}$, however, an asymmetric distribution may arise due to the contribution of the shear strain $\varepsilon_{xz}$. Such asymmetric SAW absorption of a-FMR has recently received increasing attention \cite{Mingran, Onose}. The origin of the asymmetric SAW absorption on these works was explained as interference of the longitudinal ($\varepsilon_{xx}$) and shear ($\varepsilon_{xz}$) strain components. Under this scenario one may expect that the shear strain component should be dominant or at least comparable to the longitudinal strain \cite{Onose}. However, in the thin film limit, $kd<<1$, where $k$ is the wavevector and $d$ the film thickness, the shear strain is strongly suppressed and longitudinal strain is dominant.

Even though the shear strain $\varepsilon_{xz}$ is strongly suppressed in the thin film limit, Maekawa and Tachiki theoretically demonstrated that a rotational deformation of the lattice $\upomega_{xz}=1/2(\partial u_{x}/\partial z-\partial u_{z}/\partial x)$ survives and can couple to the magnetic anisotropy (out of plane, i.e. along $z$-direction) \cite{Maekawa}, where the displacement vector components $u_{x}$ and $u_{z}$ are given by
\begin{align}
u_{x}=& Ak_{t}\left[ e^{-k_{t}z}-\frac{2q^{2}}{q^{2}+k_{t}^{2}}e^{-k_{l}z}\right] \cos(qx-\omega t) , \\
u_{z}=& -Aq\left[ e^{-k_{t}z}-\frac{2k_{t}k_{l}}{q^{2}+k_{t}^{2}}e^{-k_{l}z} \right] \sin(qx-\omega t) ,
\end{align}    
where A is the amplitude of the SAW, $k_{t}=\sqrt{q^{2}-(\omega/v_{t})^{2}}$, $k_{l}=\sqrt{q^{2}-(\omega/v_{l})^{2}}$, with $v_{t}$ and $v_{l}$ being the transverse and longitudinal sound velocities. One can compute $\upomega_{xz}$ as 
\begin{equation}
\upomega_{xz}=\frac{1}{2}Aq^{2}\upxi^{2}(e^{-k_{t}z})\cos(qx-\omega t) .
\end{equation}  
$\upxi$ is given by the ratio of the velocities $v_{t}$ and $v_{l}$. Following \cite{Maekawa}, different from the standard magnetoelastic coupling, here the rotational deformation $\upomega_{xz}$ couples to the uniaxial crystal anisotropy (out of plane, $z$) $D$ asociated with a spin $\overline{S_{i}}$ in the $i$-th site, such that 
\begin{equation}
\begin{split}
-D[S_{iz}^{2}+\upomega_{xz}(S_{iz}S_{ix}+S_{ix}S_{iz})]
\label{eq:H-NV}
\end{split}
\end{equation}
If we insert Eq.~(13) in Eq.~(14) we obtain the Hamiltonian describing the interaction between SAW and spins via the rotational deformation $\upomega_{xz}$
\begin{equation}
H=\frac{1}{2}Aq^{2}\upxi^{2}D\sum_{i}(e^{-k_{t}l_{iz}})(S_{iz}S_{ix}+S_{ix}S_{iz})\cos(ql_{ix}-\omega t)
\label{eq:Hamiltonian}
\end{equation}
where $l_{xi}$ is the $x$-component of the position vector for the $i$-th site. The interaction described by Eq.~(\ref{eq:Hamiltonian}) implies that SAW may excite surface magnons via rotational motion, and induce SAW attenuation even in the absence of shear strain. 

\begin{figure}[t!]
\begin{center}
\includegraphics[width=1.0\columnwidth, height=10.0cm, keepaspectratio]{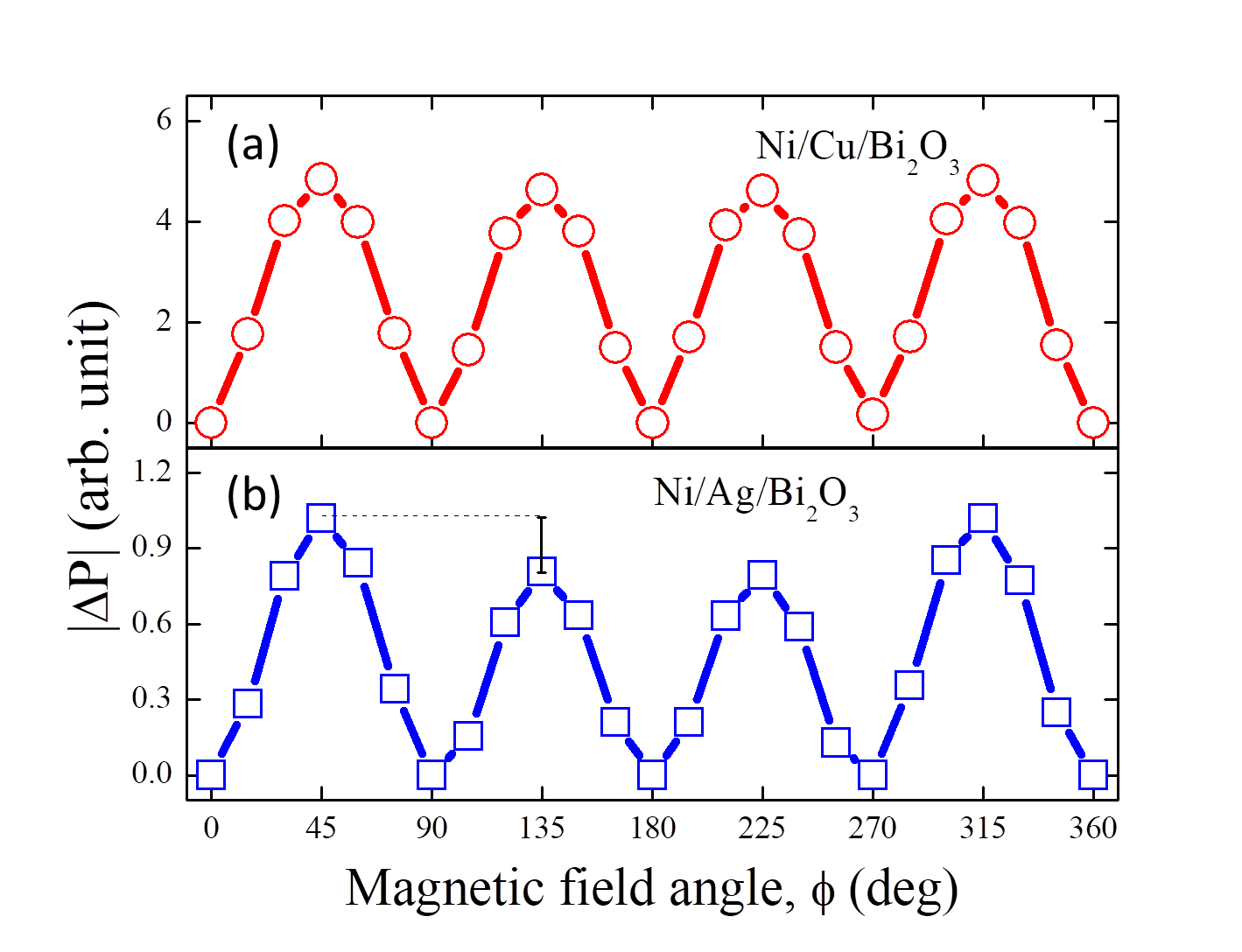}
\end{center}
\vspace*{-5mm}
\caption{Asymmetric power absorption $P_{SAW}$ of SAW in (a) Ni/Cu/Bi$_{2}$O$_{3}$ and (b) Ni/Ag/Bi$_{2}$O$_{3}$. Figures are adapted from ref.\cite{Mingran}.}
\label{fig:ABS2}
\end{figure}

Figure~\ref{fig:ABS2} shows a comparison of the in-plane magnetic field dependence of SAW attenuation for Ni/Cu/Bi$_{2}$O$_{3}$ and Ni/Ag/Bi$_{2}$O$_{3}$. Although, nominally both heterostructures contain similar Ni and Bi$_{2}$O$_{3}$, and Cu and Ag posses similar acoustic attenuation, it is possible to observe a significant difference in the asymmetric behavior of SAW absorption between the two heterostructures. In a previous report, we attributed such difference in asymmetric SAW absorption to interference of longitudinal and shear waves \cite{Mingran}, as was also suggested for other systems \cite{Onose}. However, the authors have recently become aware of a report suggesting enhancement of magnetic anisotropy energy in the Ag/Ni interface \cite{Tsay}. Such enhancement of magnetic anisotropy most likely is the result of spin reorientation due to intefacial spin orbit coupling; however, further studies are necessary to clarify it. Together with recent independent experiments \cite{paper}, this asymmetric absorption in figure~\ref{fig:ABS2}, may be due to the lattice rotation coupling with magnetization as described by Eq.~(\ref{eq:Hamiltonian}). 

\section{Acoustic spin pumping}

\begin{figure}[t!]
\begin{center}
\includegraphics[width=1.0\columnwidth, height=10.0cm, keepaspectratio]{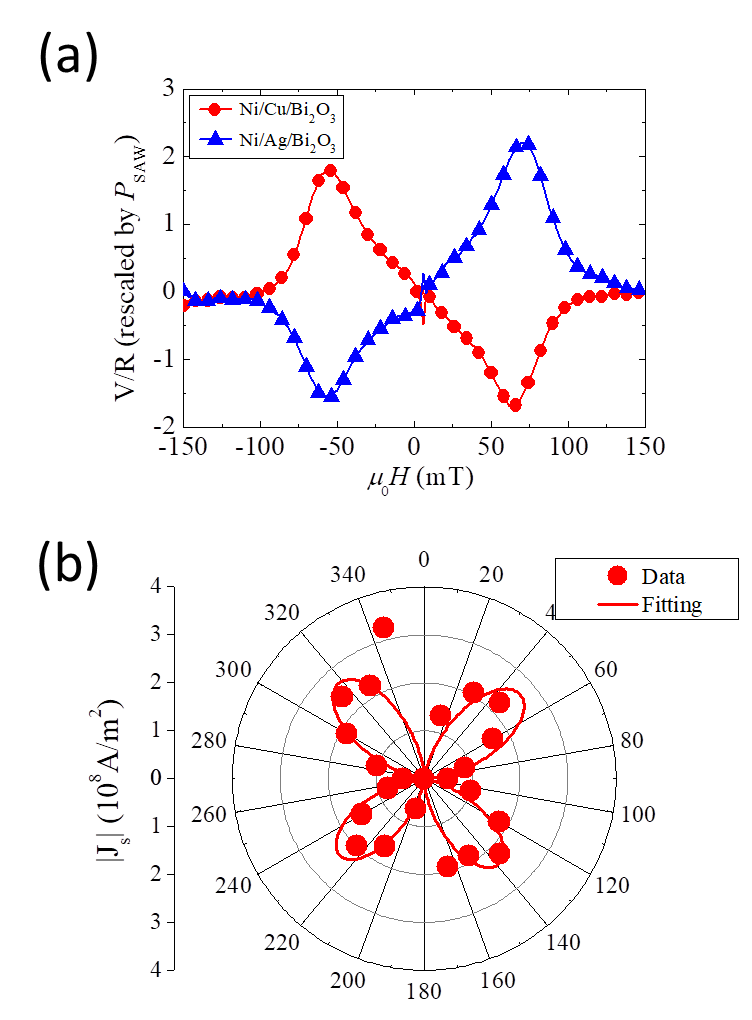}
\end{center}
\vspace*{-5mm}
\caption{(a) Inverse Edelstein effect (IEE) signal rectified at the Cu/Bi$_{2}$O$_{3}$ (red circles) and Ag/Bi$_{2}$O$_{3}$ (blue triangles) interfaces. (b) Angle dependence of the spin current density J$_{s}$ for Ni/Cu/Bi$_{2}$O$_{3}$. Figures adapted from ref.\cite{Mingran}.}
\label{fig:ABS3}
\end{figure}

\begin{figure}[t!]
\begin{center}
\includegraphics[width=1.0\columnwidth, height=10.0cm, keepaspectratio]{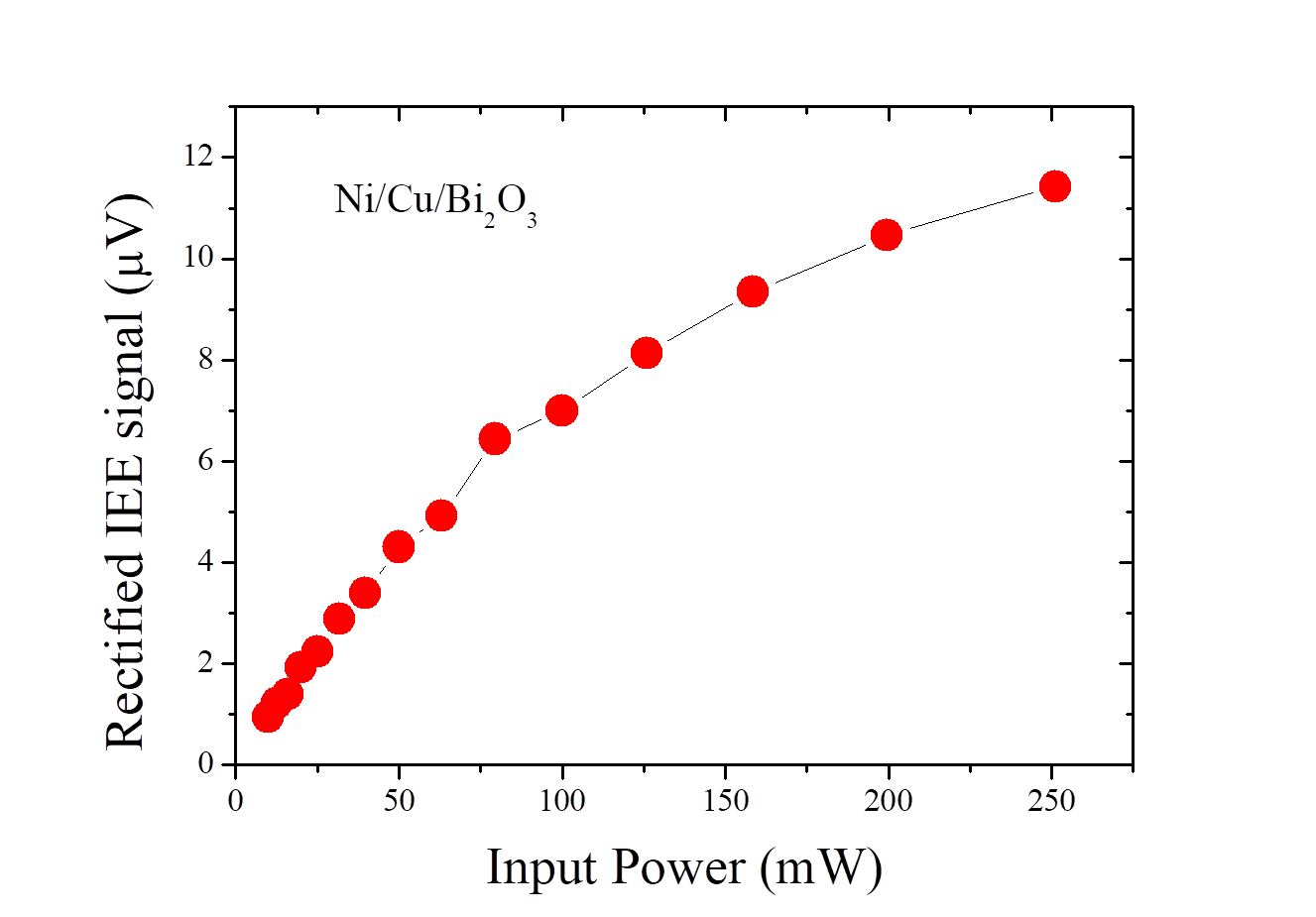}
\end{center}
\vspace*{-5mm}
\caption{Power dependence of rectified Inverse Edelstein effect (IEE) signal at the Cu/Bi$_{2}$O$_{3}$ interface}
\label{fig:ABS4}
\end{figure}

As described by Tserkovnyak et al \cite{Yaro, Yaro2}, magnetization dynamics following the LLG equation can pump spin current from a ferromagnetic layer into a nonmagnetic metal. Such spin pumping is the result of loss of torque acting in the magnetization vector, and can be directly related to enhancements in the Gilbert damping $\upalpha$. The conservation of angular momentum indicates that the damping of magnetization precession can pump angular momentum or spin current into an adjacent layer. The spin current generated by acoustic ferromagnetic resonance can be converted to charge current by either inverse spin Hall effect (ISHE) \cite{Saitoh, Wieler2} or inverse Edelstein effect (IEE) \cite{Mingran}. Figure~\ref{fig:ABS3}(a) shows the IEE signal rectified at the Cu/Bi$_{2}$O$_{3}$ (red circles) and Ag/Bi$_{2}$O$_{3}$ (blue triangles) interfaces. The opposite signs of rectified signal reflect the opposite Rashba spin splitting at these two interfaces \cite{Jorge}. We can use the IEE signal to estimate the generated spin current density $J_{s}$ with the following formula taken from \cite{Mingran}

\begin{equation}
\begin{split}\label{eq:LVolt}
J_{s}=\frac{V(\upphi)}{\uplambda_{IEE}wRsin\uptheta}
\end{split}
\end{equation}

where $V(\upphi)$ is the voltage signal detected while the magnetic field is applied at angle $\upphi$, $\uplambda_{IEE}$ is the inverse Edelstein length, $w$ is the sample width and $R$ the electric sample resistance. Figure~\ref{fig:ABS3}(b) shows the angle dependence of the spin current density for Ni/Cu/Bi$_{2}$O$_{3}$, with the following parameters: $\uplambda_{IEE}=0.17$nm, $w=10\upmu$m, $R=42.87\Omega$. The angle dependence of spin current density shows similar behavior of that of the power absorption of SAW presented in figure~\ref{fig:ABS1}(b). The spin current density is in the order of 10$^{8}$ A/m$^{2}$, same order of magnitude to spin current density generated by standard microwave photon FMR \cite{Rojas, Hoff}. For completeness, we show in figure~\ref{fig:ABS4} the power dependence of IEE signal rectified at the Cu/Bi$_{2}$O$_{3}$ interface. As the input power increases the IEE voltage increases monotonically. At low input powers the signal can be approximated to a linear increase, however, at high input power the signal increases nonlinearly.    

\section{Conclusion and outlook}

We provided an overview of the basic characteristics of SAW and description of the magnon-phonon coupling that triggers acoustic ferromagnetic resonance. We extended our discussion in a relatively yet unexplored coupling mechanism of lattice rotation and magnetic anisotropies \cite{Maekawa}. Such coupling mechanism offers a novel direction for promoting SAW asymmetric attenuation. Enhancement and modulation of asymmetric electrical charge conductivity in electronic circuits allowed the development of the electronic diode technology. The coupling mechanism between magnetic anisotropy and lattice rotation paves the way to explore significant enhancements of asymmetric SAW attenuation in the GHz frequency range, with potential for the development of the magneto-acoustic analog to an electronic diode \cite{paper}. We described and showed the spin current density generated by a-FMR, the so-called acoustic spin pumping. The order of magnitude of the spin current density J$_{s}$ is comparable to that produced by standard FMR technique. Here, acoustic wave reflectors \cite{reflectors} may represent an opportunity for further enhancement of the generated spin current densities in acoustic spin pumping.     
Beyond the specific topic presented here, SAW can also couple to nonmagnetic layers via the so-called spin rotation coupling and generate spin currents in the absence of magnetic materials or external magnetic fields \cite{Matsuo}. Initial experimental evidence of spin rotation coupling has been recently reported via spin transfer torque mechanism in Cu/NiFe bilayer \cite{Nozaki}. However, it would be interesting to demonstrate the generation of SAW induced spin current in an all-nonmagnetic structure. Coupling involving SAW has multiple applications that range from wireless technology, sensing, biology to control of elemental charges in condensed matter and coupling to quantum states of matter \cite{Roadmap}, which offers opportunities for interdisciplinary research and device developments.

\hfill \break

This work was supported by Grants-in-Aid for Scientific Research on Innovative
Areas (No. 26103001, No. 26103002) and JSPS KAKENHI (No. 19H05629). MX was supported by JSPS KAKENHI (No. JP19J21720). KY would like to acknowledge support of JSPS KAKENHI (JP 19K21040). SM was financially supported by ERATO, JST, and  KAKENHI (No 17H02927 and No. 26103006) from MEXT, Japan.


\end{document}